\def\sourcepath{\char"7E/papers/levitin/info}

\documentclass{article}
\usepackage{amsmath,amsfonts,amscd,url,fleqn}
\usepackage{enumitem}

\def\leftfrac{.35}\def\ritefrac{.65}% -*-fundamental-*-
\makeatletter
\long\def\proofbox#1{\gdef\@proofbox{#1}}
\proofbox{\small{\tt\@ifundefined{sourcepath}{}{\sourcepath/}\jobname}\\%
\ifx\UNDEF\Version\else\quad v.\Version, \fi\today}

\def\proofref#1{\proofbox{\small{\tt#1\par
	[edited by Tom Toffoli for personal use]\par
	{\tt\sourcepath/\jobname}, 
	\number\month/\number\day/\number\year
	\par}}}

 \def\affil#1{\\{\small\sl#1\par}}
 \long\def\author#1{\gdef\@author{#1}}
 \author{Tommaso Toffoli ({\tt tt\char"40bu.edu})\affil{Electrical and
Computer Engineering, Boston University, MA 02215}}

 \long\def\abstract#1{\gdef\@abstract{#1}}
 \abstract{}

\long\def\@firstoftwo#1#2{#1}
\long\def\@secondoftwo#1#2{#2}
\def\@ifundefined#1{%
  \expandafter\ifx\csname#1\endcsname\relax
    \expandafter\@firstoftwo
  \else
    \expandafter\@secondoftwo
  \fi}
 \@ifundefined{leftfrac}{\def\leftfrac{.32}}{}
 \@ifundefined{ritefrac}{\def\ritefrac{.68}}{}

\def\@maketitle{\newpage\noindent\leavevmode
  \begin{minipage}[t]{\leftfrac\textwidth}
    \hrule height0pt
    \@proofbox
  \end{minipage}\hfil
 \begin{minipage}[t]{\ritefrac\textwidth}
    \hrule height0pt
    \raggedleft
    \LARGE\@title\par
    \vskip4pt
    \large\@author
  \end{minipage}
  \vskip8pt
  \ifx\@abstract\@empty\else{\vskip.5em\leftskip1.25in\parskip4pt\small\@abstract\par\vskip.5em}\fi
  \noindent
  \rule{\textwidth}{0.4pt}
  \vskip16pt}

\makeatother

\usepackage[paper=letterpaper,dvips]{geometry} \input pack10.sty
\def\ifundefined#1{\expandafter\ifx\csname#1\endcsname\relax}

 \makeatletter
 \DeclareRobustCommand\em
        {\@nomath\em \ifdim \fontdimen\@ne\font >\z@
                       \upshape \else \slshape \fi}

\def\@begintheorem#1#2{\sl \trivlist \item[\hskip \labelsep{\bf #1\ #2}]}
\def\@opargbegintheorem#1#2#3{\sl \trivlist
     \item[\hskip \labelsep{\bf #1\ #2\ (#3)}]}

 \newenvironment{fineprint}{\par\medskip\begingroup\small}{\par\endgroup\medskip}

 \newcommand{\ie}{i.e.,~}	% id est ("that is")
 
 \newcommand{\eg}{e.g.,~}	% Exempli gratia ("for example")
 \newcommand{\cf}{cf.~}		% Confer ("see, compare with")
 \mathchardef\BY="0202
 \def\@empty{}
 \newcommand{\asin}[2][]{{%extra braces for lower level, to forget \@cite
    \def\t@mp{#1}%
    \def\@cite##1##2{\marginpar{\hfil{\footnotesize$
    \ifx\t@mp\@empty\text{##2}\else\frac{\text{##1}}{\text{##2}}\fi$}\hfil}}%
\cite[#1]{#2}}}
 \let\asin\cite

 \def\pages#1{}

 \newcommand{\sectlabel}[1]{\label{sect:#1}}
 \newcommand{\footlabel}[1]{\label{foot:#1}}
 \newcommand{\eqlabel}[1]{\label{eq:#1}}

 \newcommand{\itmlabel}[1]{\label{itm:#1}}

 \newcommand{\Chapt}[2][]{\def\t@mp{#1}%
\chapter{#2} \ifx\t@mp\@empty\else\sectlabel{#1}\fi}
 \newcommand{\Sect}[2][]{\def\t@mp{#1}%
\section{#2} \ifx\t@mp\@empty\else\sectlabel{#1}\fi}
 \newcommand{\Subsect}[2][]{\def\t@mp{#1}%
\subsection{#2} \ifx\t@mp\@empty\else\sectlabel{#1}\fi}
 \newcommand{\Subsubsect}[2][]{\def\t@mp{#1}%
\subsubsection{#2} \ifx\t@mp\@empty\else\sectlabel{#1}\fi}
 \newcommand{\Foot}[2][]{\def\t@mp{#1}%
\footnote{#2\ifx\t@mp\@empty\else\footlabel{#1}\fi}}
 \newcommand{\Item}[2][]{\def\t@mp{#1}%
\item \ifx\t@mp\@empty\else\itmlabel{#1}\fi#2}
 \newcommand{\Eq}[2][]{\def\t@mp{#1}%
\begin{equation}#2\ifx\t@mp\@empty\notag\else\eqlabel{#1}\fi\end{equation}}
 \newcommand{\Eqaligned}[2][]{\def\t@mp{#1}%
\begin{equation}\begin{aligned}#2\end{aligned}
\ifx\t@mp\@empty\notag\else\eqlabel{#1}\fi
\end{equation}}
 \newcommand{\Eqmultline}[2][]{\def\t@mp{#1}%
\begin{multline}#2\ifx\t@mp\@empty\notag\else\eqlabel{#1}\fi
\end{multline}}
 \newcommand{\Eqgathered}[2][]{\def\t@mp{#1}%
\begin{equation}\begin{gathered}#2\end{gathered}
\ifx\t@mp\@empty\notag\else\eqlabel{#1}\fi
\end{equation}}

  \newcommand{\sect}[1]{\S\ref{sect:#1}}      % Ref. to section or subsection
 \newcommand{\eq}[1]{(\ref{eq:#1})}	% Ref. to equation
 \def\thlabel#1{\label{th:#1}}

 \newcommand{\Theor}[2][]{\def\t@mp{#1}%
\begin{theorem}#2\ifx\t@mp\@empty\else\thlabel{#1}\fi\end{theorem}}
 \newcommand{\Lemma}[2][]{\def\t@mp{#1}%
\begin{lemm}#2\ifx\t@mp\@empty\else\thlabel{#1}\fi\end{lemm}}
 \newcommand{\Defi}[2][]{\def\t@mp{#1}%
\begin{definit}#2\ifx\t@mp\@empty\else\thlabel{#1}\fi\end{definit}}
 \newcommand{\Prop}[2][]{\def\t@mp{#1}%
\begin{proposition}#2\ifx\t@mp\@empty\else\thlabel{#1}\fi\end{proposition}}

 \long\def\endsubsection#1{\smallskip\hbox to\hsize{\leaders\hrule\hfill\ \sect{#1}}\medskip}

  \def\@arabic#1{\number #1} % my redefinition

\long\def\@makecaption#1#2{
	\vskip\abovecaptionskip
	\sbox\@tempboxa{{\small #1: #2}}%
	\ifdim\wd\@tempboxa>\hsize
	    {\small #1: #2\par}
	\else
	   \global\@minipagefalse
	   \hbox to\hsize{\hfil\box\@tempboxa\hfil}
	\fi
	\vskip\belowcaptionskip}

\def\figstrut#1{\hbox to\linewidth{\vrule height#1\hfill}}

\def\cstrip#1{\setbox0=\hbox{$#1$}\kern-.5\wd0\lower2pt\box0}
\def\rstrip#1{\setbox0=\hbox{$#1$}\kern-\wd0\lower2pt\box0}
\def\lstrip#1{\setbox0=\hbox{$#1$}\lower2pt\box0}
\def\tstrip#1{\setbox0=\hbox{$#1$}\kern-.5\wd0\lower\ht0\box0}
\def\bstrip#1{\setbox0=\hbox{$#1$}\kern-.5\wd0\raise\ht0\box0}
\def\Lstrip#1{\setbox0=\hbox{$\mskip2mu#1$}\lower2pt\box0}

\def\icpad{\thinspace}
\def\codepad{.3em}
\def\codeskip{\kern1pt}
\let\codestop\relax

\def\ic{\icpad\begingroup \tt \let\do\@makeother \dospecials 
          \@ifstar{\@sic}{\@ic}}
\def\@sic#1{\def\@tempa ##1#1{##1\endgroup\icpad}\@tempa}
\def\@ic{\obeyspaces \frenchspacing \@sic}

\def\codeline{\par\codeskip\noindent\begingroup\leftskip=\codepad\tt \let\do\@makeother \dospecials 
          \@ifstar{\@scodeline}{\@codeline}}
\def\@scodeline#1{\def\@tempa ##1#1{##1\ \codestop\par\endgroup\codeskip}\@tempa}
\def\@codeline{\obeyspaces \frenchspacing \@scodeline}

\def\code{\@code\frenchspacing\@vobeyspaces\verbatim@start}
\def\finecode{\par\footnotesize\@code\frenchspacing\@vobeyspaces\verbatim@start}
\def\@code{\the\every@verbatim
 \par\noindent\fussy
  \@beginparpenalty \predisplaypenalty
  \leftskip=\codepad\rightskip\z@
  \parindent\z@\parfillskip\@flushglue\parskip\z@
  \@@par
  \def\par{%
    \if@tempswa
      \leavevmode\null\@@par\penalty\interlinepenalty
    \else
      \@tempswatrue
      \ifhmode\@@par\penalty\interlinepenalty\fi
    \fi}%
  \def\@noitemerr{\@warning{No verbatim text}}%
  \obeylines
  \verbatim@font
  \let\do\@makeother \dospecials
  \everypar \expandafter{\the\everypar \unpenalty}}

\makeatother

\newcommand{\Codeline}[1][]{\def\t@mp{#1}%
\par\smallskip\noindent\begingroup\leftskip=\codepad\tt \let\do\@makeother \dospecials 
          \@ifstar{\@sCodeline}{\@Codeline}}
\def\@sCodeline#1{\def\@tempa ##1#1{##1\par\endgroup\smallskip}\@tempa}
\def\@Codeline{\obeyspaces \frenchspacing \@sCodeline}

 \title{A quantum limit on the information\\retrievable from an image}
 \author{Lev B Levitin and Tommaso Toffoli\\Boston University, ECE Dept.}
 \abstract{We consider the physical limitations imposed on the information content of an image by the wave and quantum nature of light, when the image is obtained by illuminating a reflecting or transmitting planar object by natural---i.e., fully thermalized---light, or by observation of an object emitting incoherent (thermal) radiation. The discreteness of the degrees of freedom and the statistical properties of thermal radiation are taken into account. We derive the maximum amount of information that can be retrieved from the object. This amount is always finite and is proportional to the area of the object, the solid angle under which the entrance pupil of the receiver is seen from the object, and the time of observation.

  An explicit expression for the information in the case where the information recorded by the receiver obeys Planck's spectral distribution is obtained.  The amount of information per photon of recorded radiation is a universal numerical constant, independent of the parameters of observation.}
\begin{document}
\maketitle

\def\nbar{\overline n}

\Sect{Introduction}

A very general and most common way to store information is to encode it in the
optical properties of an object. One can then retrieve information by viewing
the object by reflected or transmitted light---or even by light emitted by the
object itself---for a specified time interval. Such methods, using natural
(\ie\ thermalized) light, are universally used in photography, television,
printing, etc. In the present paper we discuss the question of what are the
fundamental limitations imposed on these methods by the wave and quantum nature
of light and by the statistical properties of thermal radiation, and we
determine the maximum amount of information that can be retrieved from the
object by illuminating it in this way and observing its optical image during a
read-out window of a prescribed time width. We exclude from consideration
cases, like holography, which make use of coherent rather than thermal
radiation.

The problem under consideration can be reduced to the evaluation of the
capacity of a specific classical-quantum information channel (a channel where
classical signals are encoded within quantum states).

Though the bulk of work in this direction was done throughout the last few
decades (\eg
\cite{davies77,gordon62,lebedev66,levitin69,levitin87,levitin94,mitiugov,schumacher97}),
the special case of thermal radiation deserves a detailed analysis. Note that
thermal radiation has maximum entropy for a given average occupation number
(the average number of photons per field oscillator); therefore, it is the most
``noisy'' information carrier.

\Sect{Formulation of the problem}

For simplicity, but without loss of generality, let the external surface of the
object be a plane of area $S$. The optical properties of this surface as far as
reflected light is concerned may be specified by a \emph{reflectivity}
$a(x,y,\nu)$---a function of a point $(x,y)$ of the surface and of the
frequency $\nu$ of the radiation.  (Here and below we take this to be the
reflectivity in the direction to the observer; it is inessential for us whether
the reflection is diffuse or specular.) As far as we are only interested in
limitations owed to the physical nature of light, we can disregard the material
structure of the surface. Even though, for any material, reflectivity is
well-defined only for \emph{areas} that are large with respect to interatomic
distances, the idealization of reflectivity as a \emph{point} function is
reasonable when the wavelengths that most significantly contribute to the
spectrum of natural light sources are large with respect to interatomic
distances.

Similarly, for transmitted light, the function $a(x,y,\nu)$ should express a
\emph{transmittance}. Finally, in the case of self-luminous surface,
$a(x,y,\nu)$ will be the ratio of the spectral intensity of the radiation at a
given point of the surface to the maximum available spectral intensity.

We also assume that a function $a(x,y,\nu)$ can be assigned independently for
the two polarization states of radiation.

\Sect{Degrees of freedom of radiation}

The set of possible functions $a(x,y,\nu)$ is restricted only by the
inequality $0\leq a(x,y,\nu)\leq1$. Obviously, an arbitrarily large amount of
information could be encoded in the object by associating with it a function of
this kind. However, the use of electromagnetic radiation for reading off
information changes the situation \emph{in principle}, and this is what this
paper is about.

Indeed, suppose the area $S$ is viewed by an optical radiation detector whose
entrance pupil is seen under a solid angle $\Omega$ from any point of that
area. Then, as is well known (see \cite{gabor}), the number of ``spatial''
degrees of freedom, or, in other words, the number of field oscillators (or
photon quantum states) differing by the direction of their wave vector (or, for
another choice of set of oscillators, by the spatial localization of
photons) is equal to
 \Eq{G_s=\frac{\nu^2\Omega S}{c^2},
 }
 where $\nu$ is the frequency of a field oscillator and $c$ the speed of light.

If the object is observed for a time interval $\tau$, then the frequency
uncertainty of the photon quantum states is $\Delta\nu=1/\tau$, and the
number of states of different frequency in a frequency interval $\delta\nu$
is equal to
 \Eq{G_f=\frac{\delta\nu}{\Delta\nu} = \tau\delta\nu.
 }
 Taking into account the two possible polarization states, we obtain, for the
total number of radiation degrees of freedom,
 \Eq[degrees]{G=\frac2{c^2}\nu^2\Omega S\tau\delta\nu
 }
 (under the assumption that $\nu^2\Omega S\tau\delta\nu/c^2\gg1$, \ie\ that the
geometric optics approximation is valid); thus, we have a \emph{finite} number
of degrees of freedom.

It makes sense to associate a reflectivity $a$ only to object details which do
not exceed the space and frequency resolution of the viewer, and thus to areas
$\Delta S=c^2/\nu^2\Omega$ and frequency intervals $\Delta\nu=1/\tau$. Thus,
with the present recording/readout method, the retrievable information is
specified by the values of a set of $G$ random variables, namely, the
reflectivities for each of the field oscillators. This amount of information
will be maximum when all those random variables are independent; in that case,
it will equal the sum of the amounts of information over all the variables.

\Sect{Statistics of radiation and information}

Even for a single degree of freedom, the amount of information would be
infinite if the reflectivity $a$ could take on an infinite number of
\emph{well-distinguishable} values. But such dream becomes untenable as soon as
the quantum nature of light and the statistical properties of thermal radiation
come into play.

As is well known (see \cite{landau}), in the radiation of thermal sources the
states of field oscillators are statistically independent, and described by a
Gibbs distribution
 \Eq[gibbs]{p(n) = \frac1{\nbar+1}\big(\frac{\nbar}{\nbar+1}\big)^n
 }
 where $n$ is the occupation number---or the number of photons in a given
quantum state---and $\nbar$ the mean occupation number.

Thus, if a source of thermal radiation is used as an information transmitter,
the expected value $\nbar$---rather than the exact number of photons $n$---is
the input signal. In other words, the signal is the value of the effective
temperature for a given degree of freedom of the electromagnetic field. 

A remarkable property of distribution \eq{gibbs} of thermal radiation is its
stability with respect to scattering, aperture constraints, losses due to small
quantum efficiency of the photodetectors, etc. In particular, if the surface is
illuminated by thermal radiation, and thus obeying Gibbs's distribution, the
reflected light also has a Gibbs distribution, with the states of field
oscillators being independent provided that the illuminating beam is rather wide
and has sufficiently large bandwidth. This follows from the transformational
properties of the states of the electromagnetic field with respect to a change
from a complete set of field oscillators to another complete set (see,
\eg \cite{mitiugov}).

\medskip

Since the density matrix of thermal radiation is diagonal in energy (or
occupation number) representation, the optimal strategy to extract maximum
information is measuring the energy, or, in other words, counting the number of
photons reaching the receiver. This follows from the fact that the radiation
quantum state is completely described by distribution \eq{gibbs}
(\cf\ \cite{levitin69}).

Let $P(\nu)$ be the average energy of a field oscillator of the radiation
source of frequency $\nu$, and let $r(\nu)$ be the fraction of the energy which
is recorded by the detector in the calibration case of reflectivity $a$ equal
to unity. Then, assigning a specific value of $a$ means specifying the mean
number $\nbar$ of recorded photons for each field oscillator, or
 \Eq{\nbar=\frac{ar(\nu)P(\nu)}{h\nu}.
 }
 The \emph{maximum} mean number of recorded photons will of course be
 \Eq{\nbar_m=\frac{r(\nu)P(\nu)}{h\nu}.
 }
 
As shown in Appendix, when the mean number of photons distributed according
to \eq{gibbs} is limited by a maximum value $\nbar_m\leq9$ (in the optical
range, this corresponds to temperatures $T\leq3\cdot{10}^5\,^\circ K$), optimal
encoding is achieved by using as signal levels just \emph{two}
mean-photon-number values, namely, $\nbar=0$ and $\nbar=\nbar_m$
(corresponding to reflectivity values $a=0$ and $a=1$).  In this case, the
maximum amount of information per field oscillator is, in nats,
 {\small
 \def\r{r(\nu)}
 \def\P{P(\nu)}
  \Eq[imax]{I_m(\nu)=
   \ln\Big[1+\frac{\r\P}{\r\P\!+\!h\nu}
  \big(\frac{h\nu}{\r\P\!+\!h\nu}\big)^\frac{h\nu}{\r\P}\Big].
 }
 }
 % (where $\r$ and $\P$ stand for $r(\nu)$ and $P(\nu)$)
 When a spectral band from $\nu_0$ to $\nu_1$, assuming $(\nu_1-\nu_0)\tau\gg1$,
with \eq{degrees} and \eq{imax} taken into account, we obtain that the
maximin amount of information that can be obtained by observation of an object illuminated by incoherent light is

 \Eq[jmax]{J_m=\frac{2\tau\Omega S}{c^2}\!\int_{\nu_1}^{\nu_2}\!\!\!d\nu^2
I_m(\nu)\nu.
 }
 Thus, the amount of information increases proportionally to the area $S$ of the
object and the time $\tau$ of observation.

\bigskip

It will be interesting to derive an explicit expression for the amount of
information in the case when the radiation recorded by the detector obeys
Planck's spectral distribution (for instance, when the illumination source is a
black body of temperature $T$ and $r(\nu)=1$), \ie\ when
 \Eq{r(\nu)P(\nu)=h\nu/\big(e^{\frac{h\nu}{kT}} -1\big),
 }
 where $k$ is Boltzmann's constant and $h$ Planck's constant.
 Let the frequency band be infinite.\Foot
 {At low frequencies, geometrical optics is not valid, and one cannot use
\eq{degrees}; furthermore, for $r(\nu)P(\nu)/h\nu>9$, \eq{imax} is not
correct. But the contribution of low frequencies is small due to the factor
$\nu^2$ in the integrand, and this allows one to extend the range of
integration to zero.}
 Then
 \Eqaligned{
 J_m&=\frac{2\tau\Omega S}{c^2} \int_0^\infty\!\!\! \nu^2 \ln
  \big[1+e^{-\frac{h\nu}{kT}}(1-e^{-\frac{h\nu}{kT}})^{e^\frac{h\nu}{kT}-1}\big]d\nu\\
  &=\frac{2\tau\Omega S(kT)^3}{c^3h^3}\sigma,
 }
 where $\sigma$ is a numerical constant, given by
 \Eq{
 \sigma= \int_0^\infty\!\!\! x^2 \ln \left(1-e^{-x}(1-e^{-x})^{e^x-1}\right)dx\approx0.772
 }
 
The total number of photons recorded by the receiver is in this case
 \Eq{
 \frac{2\tau\Omega S(kT)^3}{c^3h^3}\eta,
 }
 where $\eta$ is another constant, namely,
 \Eq{
 \eta=\int_0^\infty\frac{x^2(1-e^{-x})^{e^x-1}}{e^{x-1}+(1-e^{-x})^{e^x}}dx\approx0.909\,.
 }
 Thus, the amount of information per photon is
 \Eq{\frac{J_m}N = \frac\sigma\eta\approx0.849
 \frac{\text{nats}}{\text{photon}}\approx1.225\frac{\text{bits}}{\text{photon}}.
 }
 This is a universal constant, as it does not depend on the parameters $\tau$,
 $\Omega$, $S$ of observation.

 \medskip

Now let the solid angle $\Omega$ take on the maximum possible value $2\pi$. 
Then the maximum amount $R_m$ of information per unit time per unit area is
 \Eq{
  R_m=\frac{4\pi(kT)^3}{c^2h^3}\sigma = \frac{\sigma\sqrt2}{\pi^2\sqrt c}
  \big(\frac{15P}{\pi h}\big)^\frac34,
 }
 where
 \Eq{P= \frac{4\pi}{c^2}
   \int_0^\infty\!\!\! \nu^2 \frac{h\nu}{e^\frac{h\nu}{kT}-1}d\nu=
  \frac{4\pi^5(kT)^4}{15c^2h^3}
 }
 is the energy flux of reflected radiation at maximum signal level per unit
area.

Thus, the maximum amount of retrieved information grows with the 3/4 power of
radiation energy.

\section*{Appendix}

Let us calculate the maximum information per degree of freedom of radiation.
Denote, for convenience,
 \Eq{
 \frac\nbar{\nbar+1}=x\quad\text{and}\quad\frac{\nbar_m}{\nbar_m+1}=x_m.
 }
 For a given $x$, the distribution of the number of recorded photons is
 \Eq[dist]{
 p(n/x)=(1-x)x^n.
 }
 Let $f(x)$ the probability density function of the variable $x$. Then the
information in the number $n$ of recorded photons about the signal
value $x$ is
 \Eqaligned[info]{
   I&=\sum_{n=0}^\infty\int_0^{x_m}\!\!\!\!f(x)p(n/x)\Big[\ln p(n/x)-
 \\&\quad-\ln\!\int_0^{x_m}\!\!\!\!f(x)p(n/x)dx\Big]dx
 \\&=\int_0^{x_m}\!\!\!\!f(x)[\ln(1-x)+\frac x{1-x}\ln x]dx-
 \\&\quad-\sum_{n=0}^\infty\int_0^{x_m}\!\!\!\!f(x)(1-x)x^ndx\,\ln\!\int_0^{x_m}\!\!\!\!f(x)(1-x)x^ndx.
}
 Variational analysis of the functional $I$ (see \cite{levitin65}) brings up a
remarkable result. That is, if $x_m\leq0.9$ (or $\nbar\leq9$), the maximun
information is achieved for a distribution of the form
 \Eq[maxinfo]{
 f(x)=\big(1-\frac{x_0}{x_m}\big)\delta(x)+\frac{x_0}{x_m}\delta(x-x_m),
 }    
 where
 \Eq[maxinfo+]{
 x_0=\frac{x_m}{(1-x_m)^\frac{x_m-1}{x_m}+x_m}.
 } 
 Thus, if the average occupation numbers are not too large, it is optimal to
use only two signal levels, $0$ and $x_m$. The reason for that effect is
the breadth of distribution \eq{dist}, which makes intermediate values of the
signal poorly distinguishable.

From \eq{info}, \eq{maxinfo}, and \eq{maxinfo+} we obtain that the maximum
information per degree of freedom (one field oscillator) is
 \Eq{I_m=\ln\big[1+x_m(1-x_m)^\frac{1-x_m}{x_m}\big],
 }
 corresponding to \eq{imax}.

% small

\end{document}